\documentclass[manuscript]{aastex} 
\usepackage{graphicx}
\usepackage{graphics}
\usepackage{enumerate}
\usepackage{longtable}

\begin{document}
	
\title{Merged-beams Reaction Studies of O + H$_3^+$}
	
\author{N. de Ruette\altaffilmark{1,2}, K. A. Miller\altaffilmark{1},
  A. P. O'Connor\altaffilmark{1,3}, X. Urbain\altaffilmark{4},
  C. F. Buzard\altaffilmark{5}, S. Vissapragada\altaffilmark{1}, and
  D. W. Savin\altaffilmark{1}}

\altaffiltext{1}{Columbia Astrophysics Laboratory, Columbia
  University, New York, NY 10027, U.S.A.}

\altaffiltext{2}{Present address: Department of Physics, Stockholm
  University, Stockholm, \\ 106 91, Sweden}

\altaffiltext{3}{Present address: Max-Planck Institute for Nuclear
  Physics, Heidelberg 69117, Germany}

\altaffiltext{4}{Institute of Condensed Matter and Nanosciences,
  Universit\'e catholique de Louvain, \\ B-1348 Louvain-la-Neuve,
  Belgium}

\altaffiltext{5}{Barnard College, Columbia University, New York, 
NY 10027, U.S.A.}


\begin{abstract}

We have measured the reaction of ${\rm{O}}+{{\rm{H}}}_{3}^{+}$ forming
OH$^+$ and ${\mathrm{H_2O}}^{+}$. This is one of the 
key gas-phase astrochemical processes initiating the formation of
water molecules in dense molecular clouds. For this work, we have
used a novel merged fast-beams apparatus which overlaps a beam of
H$_3^+$ onto a beam of ground-term neutral O. Here, we present cross
section data for forming OH$^+$ and ${\mathrm{H_2O}}^{+}$ at
relative energies from $\approx 3.5$~meV to $\approx15.5$ and
0.13~eV, respectively. Measurements were performed for statistically
populated O${(}^{3}{P}_{J})$ in the ground term reacting with hot
H${}_{3}^{+}$ (with an internal temperature of $\sim2500-3000$~K).
From these data, we have derived rate coefficients for translational
temperatures from $\approx25$ K to $\sim{10}^{5}$ and ${10}^{3}$ K,
respectively. Using state-of-the-art theoretical methods as a guide,
we have converted these results to a thermal rate coefficient for
forming either OH$^+$ or H$_2$O$^+$, thereby accounting for the
temperature dependence of the O fine-structure levels.  Our results
are in good agreement with two independent flowing afterglow
measurements at a temperature of $\approx300$~K, and with a
corresponding level of H$_3^+$ internal excitation. This good
agreement strongly suggests that the internal excitation of the
H$_3^+$ does not play a significant role in this reaction. The
Langevin rate coefficient is in reasonable agreement with the
experimental results at 10~K but a factor of $\sim2$ larger at 300~K.
The two published classical trajectory studies using quantum
mechanical potential energy surfaces lie a factor of $\sim1.5$ above
our experimental results over this $10-300$~K range.
		
\end{abstract}
	

	\section{Introduction}
	
The genesis of life is believed to depend, in part, on the presence of
water. Hence, understanding interstellar formation of H$_2$O is an
important subject for astrochemistry and astrobiology
\citep{Klip10a}. Gas-phase formation of water in both diffuse and
dense molecular clouds is predicted to involve reactions of neutral O
with H$_3^+$ \citep{Smit95} via
\begin{eqnarray}
\mathrm{O} (^3P) + \mathrm{H}_3^+ (^1\mathrm{A}^\prime_1) &\rightarrow& \mathrm{OH}^+ +\mathrm{H}_2, \label{oh+}\\ 
 &\rightarrow& \mathrm{H}_2\mathrm{O}^+ + \mathrm{H}. \label{o2h+}
\end{eqnarray}
Through a series of subsequent hydrogen abstraction reactions with the
abundant H$_2$, the ionic products of Reactions (\ref{oh+}) and
(\ref{o2h+}) go on to form H$_3$O$^+$. Dissociative recombination of
H$_3$O$^+$ with electrons results in several possible neutral
products, one of which is H$_2$O. The exact branching ratio for this
specific outgoing channel was uncertain for a long time, but has
recently been measured to be $\sim$ 20\% \citep{Novotny}. At this
point, one of the largest remaining uncertainties in the kinetics of
the gas-phase formation of water appears to be the lack of reliable
rate coefficients for Reactions (\ref{oh+}) and (\ref{o2h+}).
	
These two reactions also play a role in our understanding of diffuse
and dense molecular clouds. The chemistry of these clouds is driven in
part by the cosmic ray ionization rate of H$_2$ (CRIR) $\zeta$. The
temperatures of these environments are too cold for neutral-neutral
reactions to overcome typical activation barriers. Instead, gas-phase
chemistry proceeds largely by ion-neutral reactions
\citep{Herb73,Smit95}, a process that is initiated by cosmic ray
ionization. The CRIR in diffuse clouds can be inferred using the
OH$^+$ and H$_2$O$^+$ abundances. These are affected by
Reactions~(\ref{oh+}) and (\ref{o2h+}) as described by
\citet{Hollenbach}. In dense clouds $\zeta$ is constrained using the
measured H$_3^+$ abundance, which is also affected by
Reactions~(\ref{oh+}) and (\ref{o2h+}) as is discussed in
\citet{Klip10a}. Hence, constraining the astrophysical conditions in
molecular clouds requires reliable data for these two reactions.
	
Published theoretical calculations for these reactions have been
carried out using the classical Langevin method
\citep[e.g.,][]{Mill00a}. Classical trajectory studies using quantum
mechanical potential energy surfaces (PESs) have also been published
by \citet{Bett99} and \citet{Klip10a}. These latter two semi-classical
results show reasonable agreement between one another. No fully
quantum mechanical calculations exist as such calculations for
reactions involving four or more atoms are too complex for current
capabilities \citep{Alth03a,Bowm11a}.

Our previous experimental work for the analogous reactions \citep{OCon14a}
\begin{eqnarray}
\mathrm{C}(^3P) + \mathrm{H}_3^+(^1\mathrm{A}^\prime_1)  
&\to& \mathrm{CH}^+ + \mathrm{H}_2, \label{ch+} \\ 
&\to& \rm CH_2^+ + H \label{ch2+},
\end{eqnarray}
suggests that both the classical and semi-classical methods are likely
to overestimate the rate coefficient for Reaction (\ref{oh+}),
underestimate it for Reaction (\ref{o2h+}), and incorrectly predict
the temperature dependence of each.
Unfortunately, the published room-temperature, flowing afterglow
results for Reactions (\ref{oh+}) and (\ref{o2h+}) cannot resolve this
issue due to their large error bars \citep{Mill00a} and single
temperature.  Our work here aims to improve our understanding of the
${\rm O + H_3^+}$ reaction and address some of these outstanding
issues.

The rest of this paper is organized as follows. In Section
\ref{apparatus}, we describe briefly our experimental apparatus and
method.  We present and discuss our results in Section \ref{results}
and Section \ref{discussion}, respectively. The astrochemical
implications are discussed in Section \ref{astro}. A summary is given
in Section~\ref{summary}.
	
\section{Experimental Apparatus and Method} 
\label{apparatus}

A detailed description of the ion-neutral merged-beams apparatus used
for the present results can be found in \citet{OCon14a}. Here we
provide a brief description of the experiment and method, highlighting
mainly those aspects which are new or specific to the present study.

\subsection{General} 
\label{general}
	
Using a Cs-ion sputter source followed by a Wien filter, we formed a
$^{16}$O$^-$ beam at a kinetic energy of $E_{\rm O} \approx 28$~keV
($\approx 1.75$~keV/amu). This generated a pure beam of O$^-$($^2P$)
as the oxygen anion possesses only a single bound term
\citep{Rienstra}. The target material used in the source was
Al$_2$O$_3$.
	
	The neutral O beam was produced via photodetachment of the anion inside a floating cell. The kinetic energy of the resulting O beam was controlled by varying the floating cell voltage $U_{\mathrm f}$. The cross section for photodetachment of O$^-$ has been measured by \citet{Lee79}. The electron affinity of oxygen is 1.461~eV \citep{Rienstra}. Using our 808-nm (1.53-eV) laser with $\approx  1.8$~kW of power, we estimated that approximately 2\% of the anions were converted to ground term O($^3P$). The photon energy and number density were both too low to detach into higher terms. 
	
\begin{figure}[!p]
\centering
\includegraphics[width=1\textwidth]{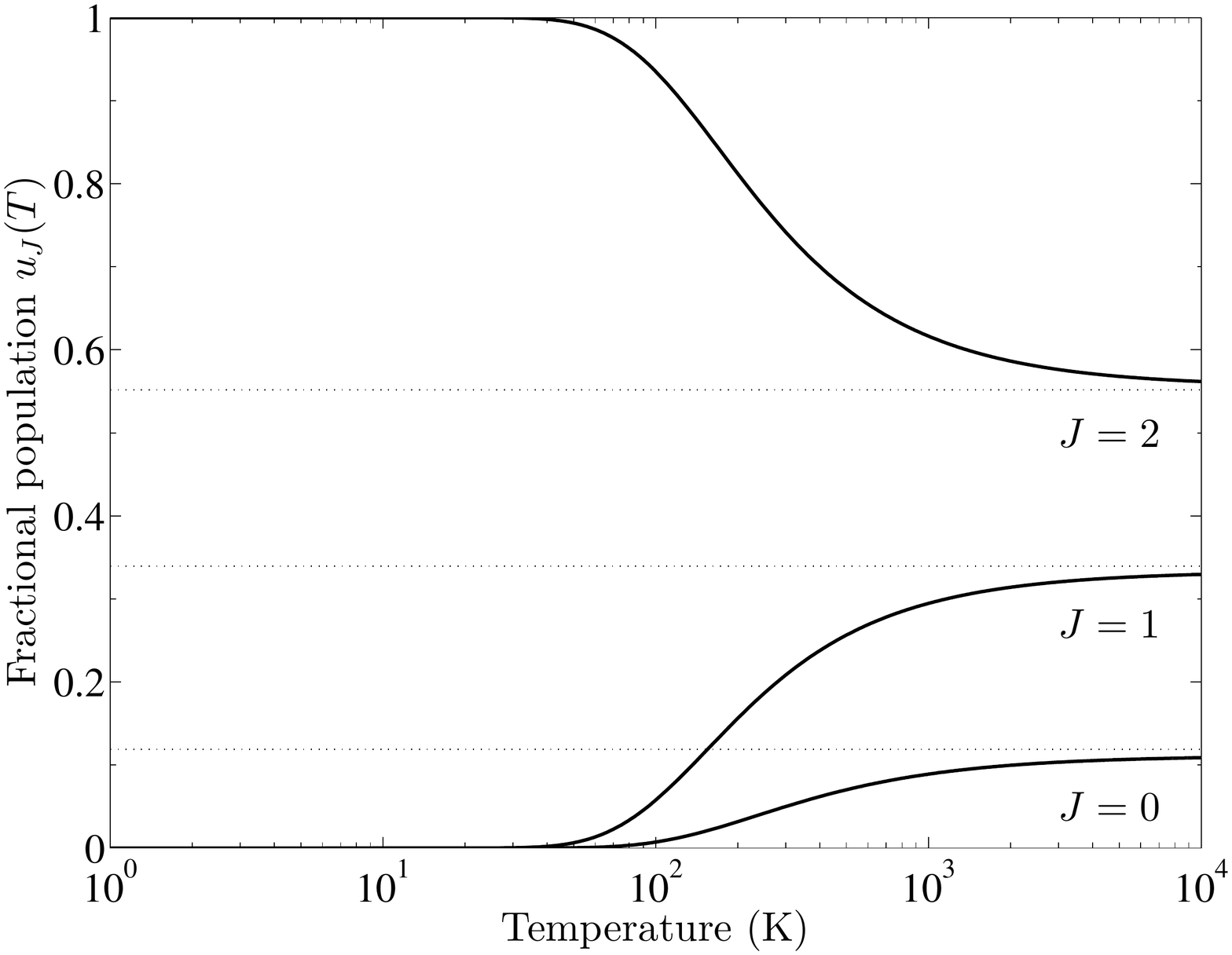}
\caption{O($^3P_J$) levels fractional population for a thermal distribution (solid curves) and a statistical distribution (dotted lines).}
\label{Fig:Olevels}
\end{figure}

Similar to the work of \citet{Sche98a} on photodetachment of C$^-$, we
expect to statistically populate all three levels in the ground term
of the atomic oxygen. The $J = 1$ and 0 fine-structure levels lie
above the $J = 2$ ground level by 19.6~meV and 28.1 meV,
respectively. The thermal population of the $J$ levels can be
calculated using the partition functions
\begin{equation}
u_J 
= 
\frac{g_J e^{(-E_J/k_{\mathrm B} T)}}{\Sigma_{J'}g_{J'} e^{(-E_{J'} /k_{\mathrm B} T)}}. 
\label{Jpartfunction}	
\end{equation} 
Here $g_J = 2J + 1$ is the statistical weight of level $J$,
$k_{\mathrm B}$ is the Boltzmann constant, and $T$ is the
temperature. These populations are presented in Figure
\ref{Fig:Olevels} where one can see that they become statistical at
temperatures above $\sim$ 1000~K. In Section \ref{sec:thermal} we
explain how we use our results with statistically populated
ground-level O to generate thermal rate coefficients.

	The molecular beamline begins with a duoplasmatron source from which we extracted a beam of cations and used a Wien filter to charge-to-mass select for H$_3^+$. The beam energy of $E_{\mathrm{H}_3^+} \approx$ 5.29 keV ($\approx 1.75$ keV/amu) was chosen to match the velocity of the oxygen anion beam.

The formation mechanism of H$_3^+$ in a duoplasmatron leads to
substantial internal excitation. In our earlier work of C on H$_3^+$
\citep{OCon14a}, we inferred an internal temperature of $\sim
2500$~K. However, we found good agreement between our thermal rate
coefficient results at 1000~K with the mass-scaled results of
\citet{Savi05a}, who studied C on D$_3^+$. Since the work of
\citet{Savi05a} used D$_3^+$ with an internal temperature of 77~K, the
good agreement between their results and those of \citet{OCon14a}
implies that the internal excitation of the H$_3^+$ does not
significantly influence reactions of the type ${\rm X + H_3^+ \to
  XH}_n^+ + \mathrm{H}_{3-n}$ for $n$ = 1 or 2.  We expect that this
will also be the case for O on H$_3^+$ and will return to this issue
in Section \ref{sec:thermal2}.

	In order to improve the H$_3^+$ beam quality in the interaction region, we have modified the beamline just before the cylindrical deflector used to merge the cation beam onto the neutral beam. In specific, we installed a set of XY steerers and adjusted the location of the one-dimensional (1D) electrostatic lens before the merger. To compensate for the lack of vertical focusing in the cylindrical deflector, we used one pair of vertical steerer plates for focusing. One of the horizontal steering plates was used to adjust the angle of the beam going into the 1D lens and cylindrical deflector. All of the other steerer plates were grounded. Additionally, the 1D lens was moved closer to the cylindrical deflector so that we could focus the beam at the appropriate location inside the deflector in order to generate a parallel beam at the exit (i.e., at the beginning of the interaction region). The resulting H$_3^+$ current in the interaction region was typically $\sim 225$~nA, corresponding to a typical number density of $\sim 10^5$~cm$^{-3}$.  

In the interaction region, the kinetic energy of any product ions
formed was essentially the sum of $E_\mathrm{O}$ plus the product of
the H$_3^+$ kinetic energy per amu ($\approx$ 1.75 keV/amu) times the
mass in amu transferred from the H$_3^+$. For O at 28 keV, the kinetic
energy of the product was 29.75 keV for forming OH$^+$, Reaction
(\ref{oh+}), and 31.5 keV for H$_2$O$^+$, Reaction (\ref{o2h+}). At
the end of the interaction region, the beams continued into a chicane
which directed the H$_3^+$ into a Faraday cup where the current was
recorded. The daughter products were directed through the chicane and
continue into an electrostatic energy analyzer. This consisted of a
series of three 90$^{\circ}$ cylindrical deflectors with voltages
optimized to direct the desired product ions into a channel electron
multiplier (CEM). This final analyzer also served to discriminate
against the dominant charged particle background which was due to the
$\sim 28$~keV O$^+$ formed by stripping of the O beam on the residual
gas downstream of the cation beam merger. The transmittance from the
interaction region to the CEM was measured to be $T_{\mathrm a}=0.74
\pm 0.02$. Here and throughout all uncertainties are quoted at an
estimated $1 \sigma$ statistical confidence level.
	
\subsection{Neutral Detector}

The neutral beam current $I_\mathrm{O}$ is monitored by measuring the
secondary negative particle emission from a target inside a neutral
particle detector \citep{Bruh10b} and is given by
\begin{equation}
I_\mathrm{O} = \frac{I_{\rm ND}}{\gamma T_{\mathrm n}}.
\end{equation}
Here $I_{\rm ND}$ is the current measured on the neutral detector,
$\gamma$ is the secondary negative particle emission coefficient, and
$T_{ n}$ is the transmission into the detector. Typical values for
$I_\mathrm{O}$ are $\sim23$~nA, as measured in amperes, with a
statistical-like uncertainty of 5\%. These currents correspond to
particle number densities of $\sim$ 10$^4$~cm$^{-3}$.
	
We determined $\gamma$ using collisional detachment of O$^-$ on helium
introduced into the chicane by a leak valve. This converts a portion
of the initial O$^-$ beam to O and O$^+$. The positive and negative
currents, $I_{\rm O^-}$ and $I_{\rm O^+}$, respectively, were measured
in a Faraday cup, called the upper cup, which is situated behind a
hole in the outer plate of the middle cylindrical deflector of the
final analyzer. Because of the design of the final analyzer we can
measure either $I_{\rm O^-}$ or $I_{\rm O^+}$ in the upper cup
simultaneous with $I_{\rm ND}$, but not all three together. The
transmittance from the interaction region into the upper cup was
measured to be $T_{\mathrm u} = 0.64 \pm 0.04$.
		
The total particle flux is assumed to be conserved at any given helium
pressure $p$. This gives
\begin{equation}
I_\mathrm{O^-}(p=0)+ I_\mathrm{O}(p=0) + I_\mathrm{O^+}(p=0) 
= 
I_\mathrm{O^-}(p) + I_\mathrm{O}(p) + I_\mathrm{O^+}(p), \label{fluxcons}
\end{equation}
where collisional detachment on the residual gas in the system
generates non-zero currents of O and O$^+$ at $p=0$. All of the
currents are taken as positive quantities. In terms of measured
quantities, we can rewrite Equation (\ref{fluxcons}) as
\begin{equation}
\frac{I_\mathrm{O^-}^{\mathrm u}(0)}{T_{\mathrm u}} + 
\frac{I_{\rm ND}(0)}{\gamma T_{\mathrm n}} + 
\frac{I_\mathrm{O^+}^{\mathrm u}(0)}{T_{\mathrm u}}  
= 
\frac{I_\mathrm{O^-}^{\mathrm u}(p)}{T_{\mathrm u}} + 
\frac{I_{\rm ND}(p)}{\gamma T_{\mathrm n}} + 
\frac{I_\mathrm{O^+}^{\mathrm u}(p)}{T_{\mathrm u}},
\end{equation}
where $I^{\mathrm u}$ stands for the currents measured in the upper
cup. Rearranging this yields
\begin{equation}
\gamma =  
\frac{-\Delta I_{\rm ND}(p)}{\Delta I_\mathrm{O^-}^{\mathrm u}(p) + \Delta I_\mathrm{O^+}^{\mathrm u}(p)} 
\frac{T_{\mathrm u}}{T_{\mathrm n}} \label{gamma}
\end{equation}
where $\Delta I(p) = I(p)-I(0)$.  Typically we measure $\Delta
I^{\mathrm u}_{\rm O^-}$ and $\Delta I_{\rm ND}$ at set of a pressures
$p_i$.  Due to the coarse control on our leak valve, we measure
$\Delta I^{\mathrm u}_{\rm O^+}$ and $\Delta I_{\rm ND}$ at a slightly
different set of pressures $p_k$.  Here we were able to match the
pressures $p_i$ and $p_k$ to better than 3\%.  This small difference
introduces an insignificant uncertainty into our $\gamma$
determination.  So we can convert these $p_k$ results to those for 
$p_i$ using
\begin{equation}
\Delta I_\mathrm{O^+}^{\mathrm u}(p_i) = 
\Delta I_\mathrm{O^+}^{\mathrm u}(p_k) 
\frac{\Delta I_{\rm ND}(p_i)}{\Delta I_{\rm ND}(p_k)} 
\end{equation}
which we can then substitute into Equation (\ref{gamma}). 

	To determine $\gamma$, we measured $I_\mathrm{O^-}^{\mathrm u}$ and $I_{\rm ND}$ simultaneously over intervals of $\sim 250$ s. First, no gas was introduced in the chicane ($p=0$). Next, we measured with a helium pressure of $p_i$. Lastly, we measured again without helium. For the currents measured without gas, we used the average of the before and after measurements. This allows us to take into account the fluctuations of the O$^-$ beam during the measurement. For each current measured at high pressure, we took the average of the current over the 250 s interval. We repeated this pattern to measure $I_{\rm ND}$ and $I_\mathrm{O^+}^{\mathrm u}$ at $p_k$. Over the several months of our $\rm O + H_3^+$ measurement campaign, we monitored $\gamma$ periodically and found $\gamma = 2.6 \pm 0.3$. The uncertainty in $\gamma$ is treated as a systematic error. 
	
\subsection{Beam Overlap and Relative Energies} 
\label{Er}

The overlap integral of the two beams has been determined using a
combination of beam profile measurements in the interaction region and
geometric modeling, as described in \citet{Bruh10a} and
\citet{OCon14a}. A typical average bulk misalignment between the two
beams of 0.38 mrad was determined from the measured beam
profiles. This represents a factor-of-two reduction compared to our
$\mathrm{C + H_3^+}$ work and is a result of the modifications of the
cation beamline described in Section~\ref{general}.
	
\begin{figure}[!t]
\begin{center}
\includegraphics[width=1\textwidth]{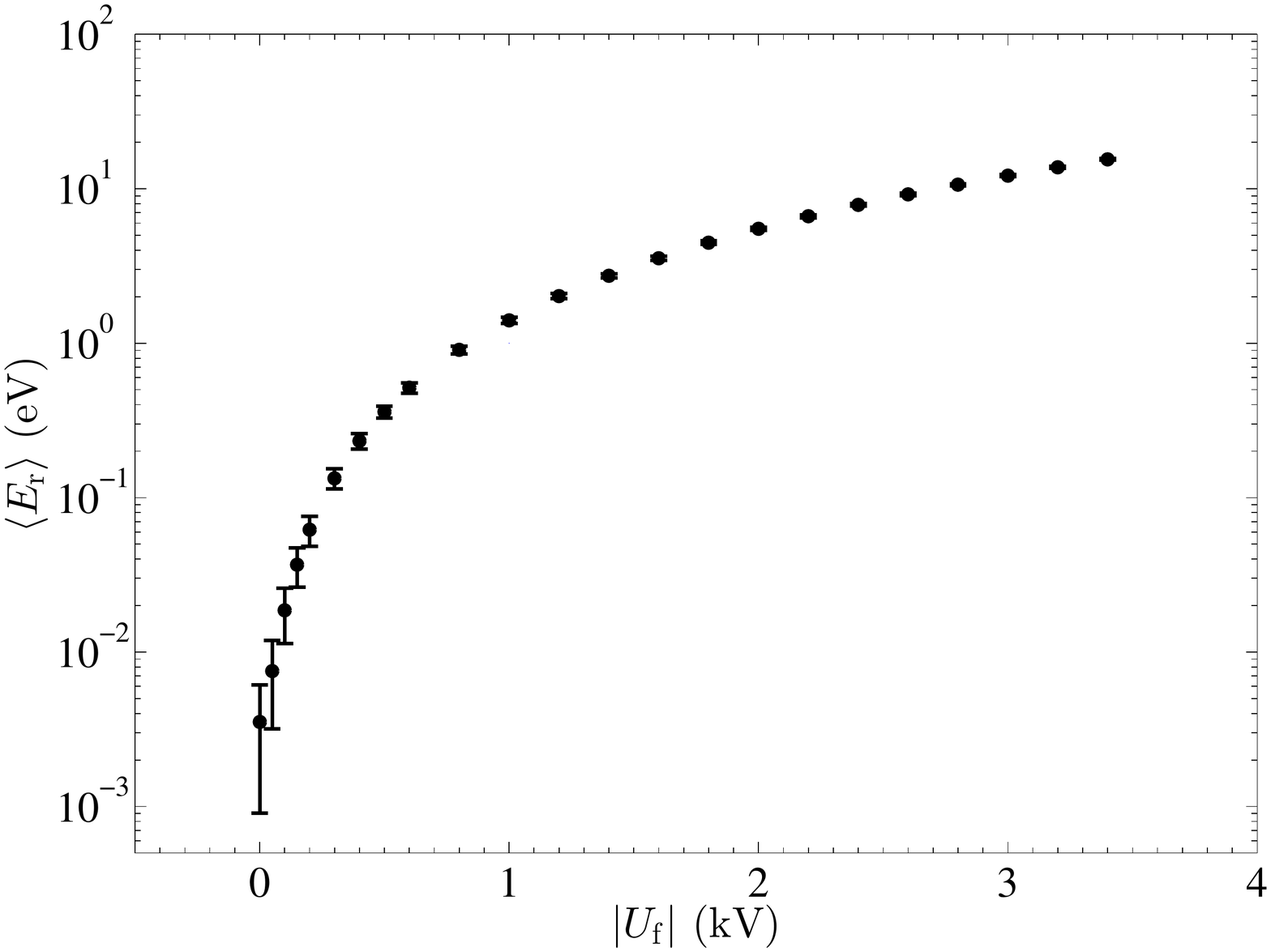}
\caption{Simulated average relative energy $\langle E_{\mathrm r}
  \rangle$ as a function of the floating cell voltage $|U_{\rm
    f}|$. The vertical error bars show the full width at half maximum
  spread of the calculated energy distribution. }
\label{Fig:energy}
\end{center}
\end{figure}

The average relative energy $\langle E_{\mathrm r} \rangle$, along
with the corresponding energy spread, was determined from the beam
profiles measurements and a Monte-Carlo simulation of the beams
trajectories described in \citet{OCon14a}. Figure~\ref{Fig:energy}
shows the results values for $\langle E_{\mathrm r}\rangle$ and the
corresponding uncertainty. Based on these simulations, the lowest
average relative energy achieved here is $\approx 3.5$ meV,
corresponding to an effective translational temperature of $\approx
27$~K (as derived from a Maxwell-Boltzmann fit of the velocity
distribution). We also find a Gaussian distribution for the
interaction angle with a mean value of $\langle \theta \rangle= 0.71
\pm 0.34$ mrad. These improvements over \citet{OCon14a} are a direct
result of the reduced bulk misalignment between the two beams.

\subsection{Merged-beams Rate Coefficient} 
\label{exprate}
	
The measured merged-beams rate coefficient is the product of the cross
section $\sigma$ and the relative velocity $v_{\mathrm r}$ convolved
with the energy spread of the experiment and is given by
\begin{equation}
\langle \sigma v_{\mathrm r} \rangle 
= 
\frac{S}{T_{\mathrm a}T_{\mathrm g}\eta} 
\frac{e^2v_{\mathrm n}v_{\mathrm i}}{I_{\mathrm n}I_{\mathrm i}} 
\frac{1}{L\langle \Omega (z) \rangle.} \label{rate} \label{rateeq}
\end{equation}
Here $S$ is the signal count rate; $T_{\mathrm g}$ is the geometrical
transmittance of the grid in front of the CEM; $\eta$ is the
efficiency of the CEM; $e$ is the elementary charge; $v_{\mathrm n}$
and $v_{\mathrm i}$ are the laboratory velocities of the neutral and
molecular ion beams, respectively; $I_{\mathrm n}$ and $I_{\mathrm i}$
are the neutral and ion beam currents, respectively; $L$ is the
interaction region length; and $\langle \Omega (z) \rangle$ is the
overlap integral of the two beams in the interaction region. Typical
values of all these parameters and their uncertainties are listed in
Tables \ref{rateval} and \ref{ratevalsys}. Details about the
associated uncertainties as well as the data acquisition procedure can
be found in \citet{OCon14a}.

\section{Experimental Results} 
\label{results}

\begin{figure}[!t]
\begin{center}
\includegraphics[width=1\textwidth]{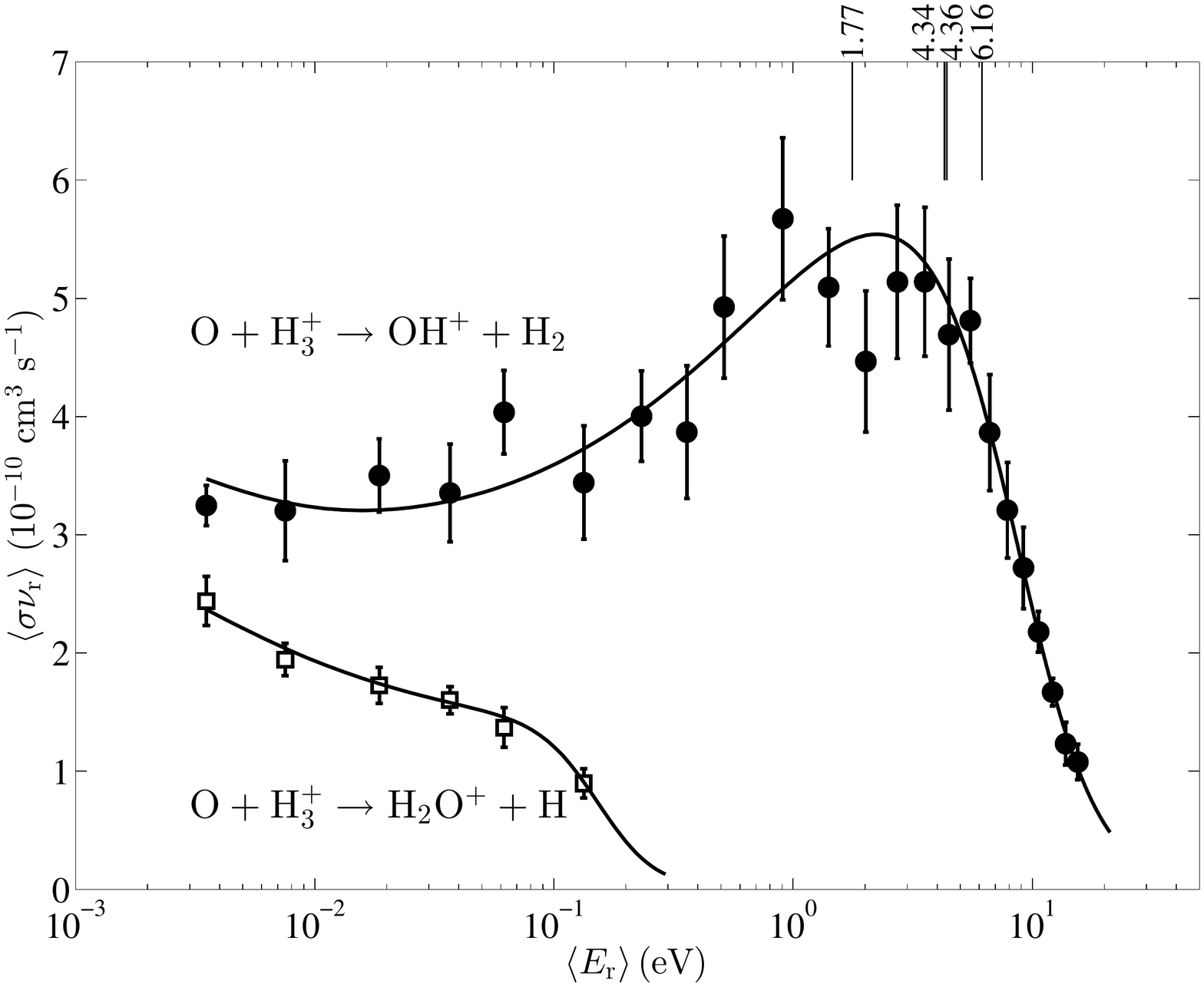} 
\caption{Experimental rate coefficients $\langle \sigma v_{\mathrm r}
  \rangle$ as a function of the average relative energy $\langle
  E_{\mathrm r} \rangle$ are shown for Reaction~(\ref{oh+}) by the
  filled circles and for Reaction~(\ref{o2h+}) by the open
  squares. The error bars signify the 1$\sigma$ statistical
  uncertainties. Empirical fits to the experimental data using
  Equation~(\ref{Eqn:CS}) are shown by the solid lines. The vertical
  lines at the top of the graph show the threshold energies for the
  competing Reactions~(\ref{competition1})-(\ref{competition2}) of
  1.77, 4.34, 4.36 and 6.16 eV, respectively.}
\label{Fig:ExpRate}
\end{center}
\end{figure}

Our results for the merged-beams rate coefficient are presented in
Figure~\ref{Fig:ExpRate} as a function of $\langle E_{\mathrm r}
\rangle$ for Reactions~(\ref{oh+}) and (\ref{o2h+}).  Data were
collected for merged-beam rate coefficient values down to $\sim 1
\times 10^{-10}$~cm$^3$~s$^{-1}$, below which the decreasing
signal-to-noise ratio made the required data acquisition times
prohibitively long.  From these, and using our calculated experimental
energy spread, we can extract the cross section for each reaction for
statistically populated ground-term O. We then generate translational
temperature rate coefficients for each reaction by convolving the
extracted cross section with a Maxwell-Boltzmann distribution
describing the reaction center-of-mass velocity distribution. However,
the internal energies of the reactants are not in thermal
equilibrium. In Section \ref{sec:thermal} we describe how we can
convert our translational temperature results into a thermal rate
coefficient relevant for astrochemistry where the internal energies of
the reactants are typically in thermal equilibrium.

\subsection{Cross Sections} 
\label{Sec.crosssection}

The cross section $\sigma_x$ can be extracted from our data using the
fitting function
\begin{equation}
\sigma_{x} = \frac{a_0+a_{1/2}E^{1/2}}{E^{2/3}+b_1E+b_2E^2+b_4E^4}.
\label{Eqn:CS}
\end{equation}
Here $x$ refers to either Reaction (\ref{oh+}) or (\ref{o2h+}),
$\sigma_{x}$ is in units of cm$^2$, and $E$ is in eV. This function
includes a term with an $E^{-2/3}$ behavior at low energies that was
chosen to match the predicted $T^{-1/6}$ dependence in the thermal
rate coefficient at low temperatures due to the charge-quadrupole
interaction \citep{Klip10a}. The other powers of $E$ have been
arbitrarily chosen to match the higher energy dependence in the
measured merged-beams rate coefficients.
	
\begin{figure}[!t]
	\begin{center}
	\includegraphics[width=1\textwidth]{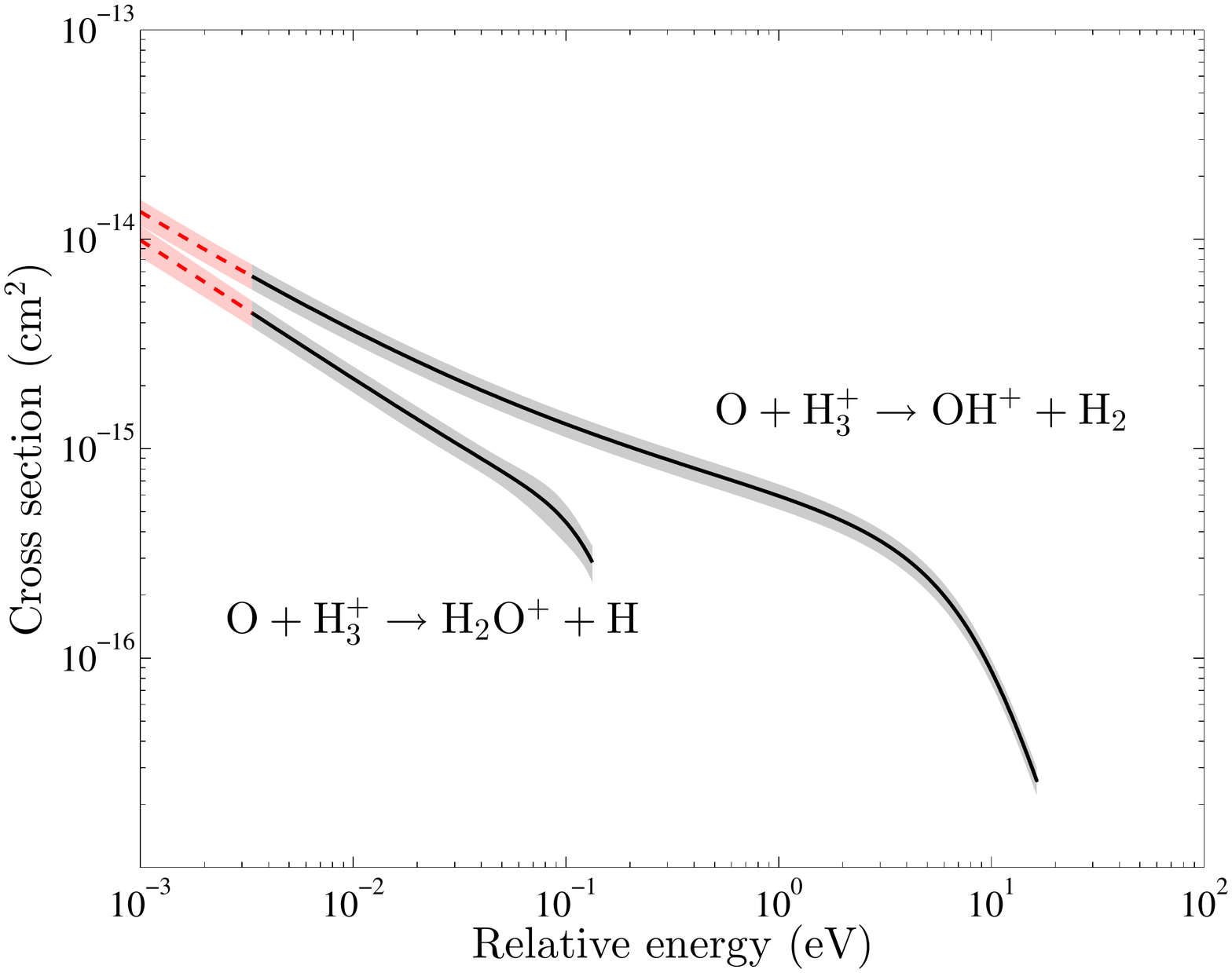}
	\caption{The solid black lines show the experimentally derived cross section $\sigma$ as a function of the average relative energy $\langle E_{\mathrm r} \rangle$ for Reactions~(\ref{oh+}) and (\ref{o2h+}). The quadrature sum of the 13\% systematic uncertainty and the fitting accuracy is denoted by the shaded region. The extrapolation of the experimental results to lower kinetic energies is shown by the red dashed lines and the systematic uncertainty, shown by the surrounding shaded region, is taken as constant using that of the lowest measured energy. 
	}
	\label{Fig:CS}
	\end{center}
	\end{figure}

To fit our measured data, we multiplied Equation~(\ref{Eqn:CS}) by
$v_{\mathrm r}$ and convolved the product with the experimental
relative velocity distribution. The best fits to the data are shown in
Figure~\ref{Fig:ExpRate} by the solid lines. Tables~\ref{tab:CS} gives
the best fit parameters of the cross section for each reaction. The
accuracy of the fit for Reaction~(\ref{oh+}) is better than 7\%. For
Reaction~(\ref{o2h+}), the accuracy is better than
15\%. Figure~\ref{Fig:CS} shows the experimentally derived cross
sections for each reaction. The results have been extrapolated to
kinetic energies below $\approx 3.5$~meV using the theory of
\citet{Klip10a} as a guide.

\subsection {Translational Temperature Rate Coefficients}

We have derived the translational temperature rate coefficient
$\alpha_x$ for each reaction using the product of the extracted cross
section $\sigma_x$ times the relative velocity, all then convolved
with a Maxwell-Boltzmann energy distribution. We fit the resulting
translational temperature rate coefficient using
\begin{equation}
\alpha_{x} = 
\frac{a_0+a_{1/2}T^{1/2}+a_1T}{T^{1/6}+b_{1/2}T^{1/2}+b_1T+b_{3/2}T^{3/2}}
\label{Eqn.thermalrate} 
\end{equation}
where $x$ refers to either Reaction~(\ref{oh+}) and (\ref{o2h+}),
$\alpha_x$ is given in units of cm$^3$~s$^{-1}$, and $T$ in units of
K. Table~\ref{tab:thermal} gives the best fit parameters for each
reaction.

\begin{figure}[!t]
	\begin{center}
	\includegraphics[width=1\textwidth]{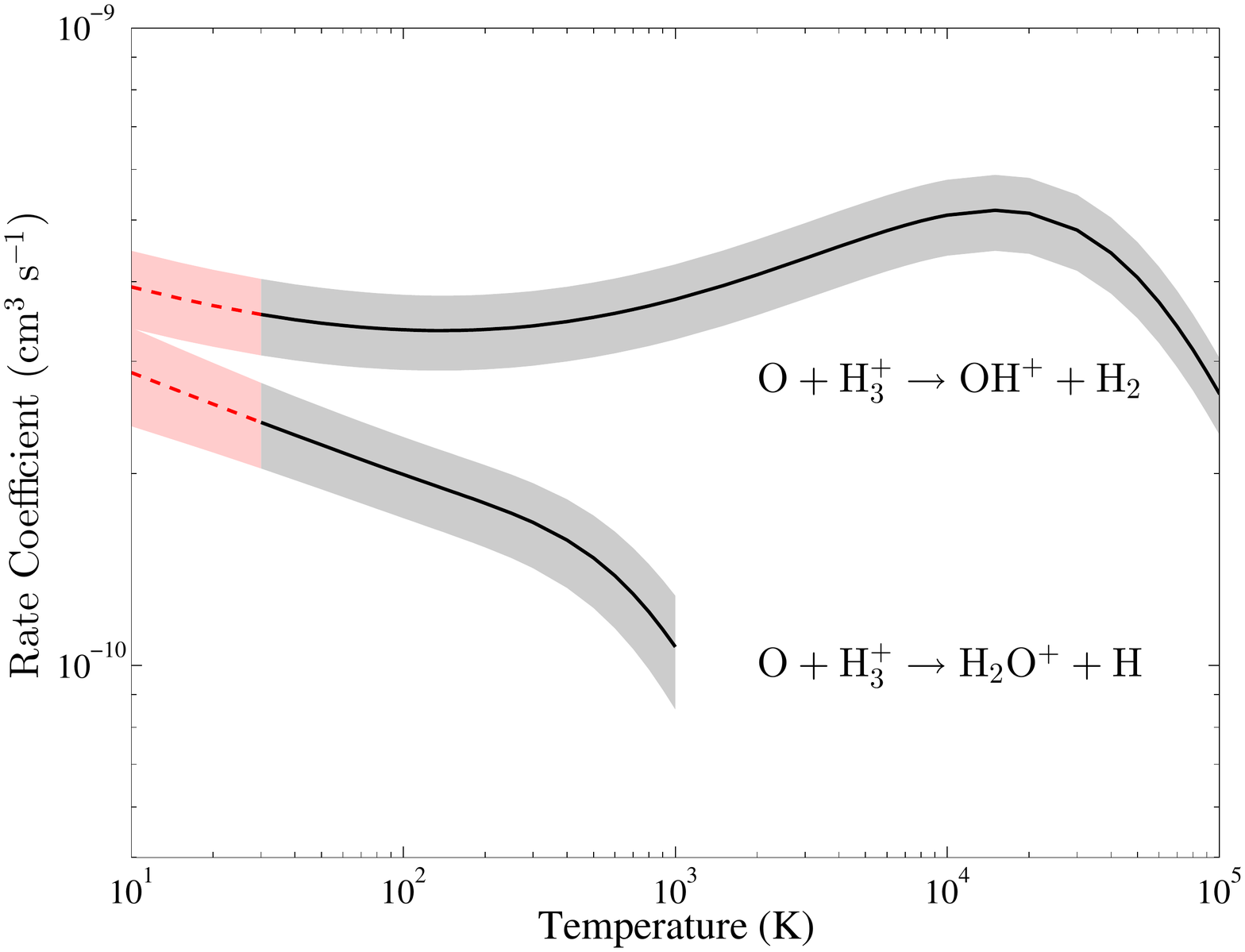}
	\caption{Experimentally derived translational temperature rate coefficients vs.\ temperature for Reactions~(\ref{oh+}) and (\ref{o2h+}) are shown by the black solid lines. The quadrature sum of the systematic uncertainty and the fitting accuracy is denoted by the shaded region. The red dashed lines are extrapolations to lower temperatures with the systematic uncertainty (shaded region) taken as constant using that of the lowest measured energy. 
	}
	\label{Fig:thermalrate}
	\end{center}
	\end{figure}

Figure~\ref{Fig:thermalrate} shows the experimentally derived
translational temperature rate coefficients and associated
uncertainties for Reactions (\ref{oh+}) and (\ref{o2h+}). The
uncertainty for each reaction, derived by adding in quadrature the
estimated systematic error and the fitting error, is $\approx 14\%$
for Reaction~(\ref{oh+}) and ranges from $\approx 15-27\%$ for
Reaction~(\ref{o2h+}), for the respective energy ranges shown in
Figure~\ref{Fig:thermalrate}. The minimum value achieved for $\langle
E_{\mathrm r} \rangle$ corresponds to an effective translational
temperature of $\approx 27$ K, as discussed earlier. The fit function
has been chosen so that the extrapolation of the translational
temperature rate coefficient below 27~K goes to a $T^{-1/6}$
dependence as predicted by \citet{Klip10a}. The highest values of
$\langle E_{\mathrm r} \rangle$ measured for Reactions~(\ref{oh+}) and
(\ref{o2h+}) of $\approx 15.5$ and 0.13~eV, respectively, yield
approximate high temperature limits of $\sim 10^5$ and $10^3$~K for
the derived translational temperature rate coefficients.

\section{Discussion} 
\label{discussion}

\begin{figure}[!t]
\begin{center}
\includegraphics[width=1\textwidth]{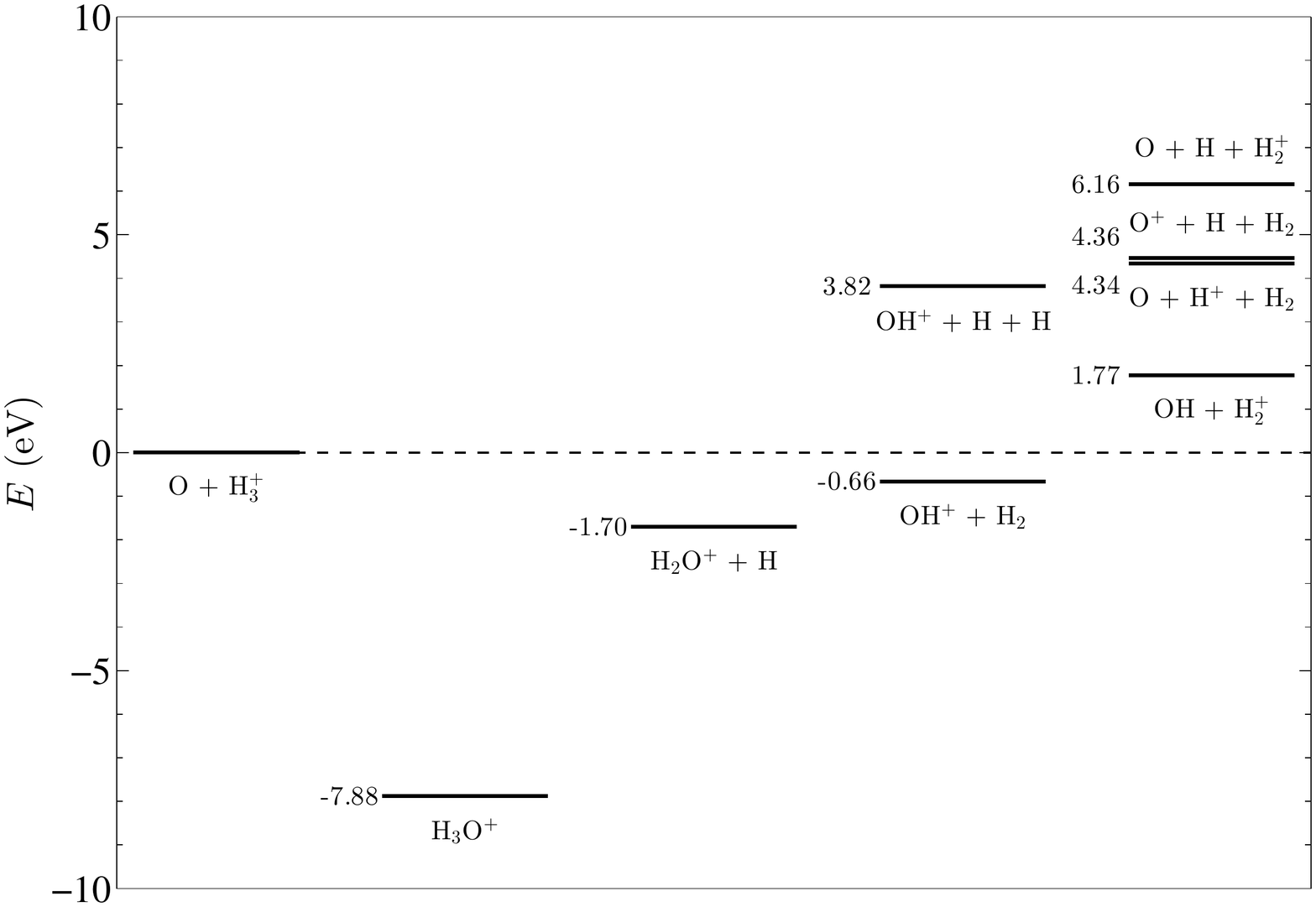}
\caption{Energy level diagram for various O + H$_3^+$ reaction
  pathways, given in eV, for the various systems in their ground
  state. Values for H$_3$O$^+$, H$_2$O$^+$ + H and OH$^+$ + H$_2$ are
  given by \citet{Mill00a}; the others are derived from \citet{Huber},
  \citet{Rohs94a}, \citet{liu2009} and \citet{NIST_ASD}.}
\label{Fig:channel}
\end{center}
\end{figure}

Figure \ref{Fig:channel} shows the asymptotic energy limits of various
O + H$_3^+$ reaction pathways. Reaction~(\ref{oh+}) is exoergic by
$\approx 0.66$ eV and Reaction (\ref{o2h+}), by $\approx 1.70$ eV
\citep{Mill00a}. The energies are given for all parent and daughter
products in their ground symmetries.
	
	\subsection{Merged-beams Rate Coefficients}

	\subsubsection{\em O + H$_3^+ \rightarrow$ OH$^+$ + H$_2$ }
	
	Taking into account the electronic states and spin symmetries of the reactants and products \citep{Bett99}, we can re-write Reaction (\ref{oh+}) as
	\begin{eqnarray}
	{\rm O}(^3P) + {\rm H}_3^+(^1{\rm A}') & \rightarrow & {\rm OH}^+(^3\Sigma^-) + {\rm H}_2(^1\Sigma_g^+).
	\end{eqnarray}
	OH$^+$ can also be formed via the endoergic reaction:
	\begin{eqnarray}
	{\rm O}(^3P) + {\rm H}_3^+(^1{\rm A}') & \rightarrow & {\rm OH}^+(^3\Sigma^-) + {\rm H}(^2S) + {\rm H}(^2S) - 3.82 \mathrm{eV}. \label{OH+other}
	\end{eqnarray}
	
The energy dependence of our measured merged-beams rate coefficient
for Reaction~(\ref{oh+}) has a somewhat similar behavior to that
measured for Reaction~(\ref{ch+}) seen by \citet{OCon14a}. The
merged-beams rate coefficient starts by increasing as the relative
energy grows. This behavior could be due to the increasing number of
ro-vibrational channels becoming energetically accessible in the
reaction products or to the opening up of new electronic states in the
intermediate reaction complex. Then, starting at about $\approx
0.9$~eV, the energy of the highest value of the merged-beams rate
coefficient, there is a suggestion of a complex structure which was
not seen in \citet{OCon14a}.  (All of the structures discussed here
appears at lower energies than expected, by about 0.9~eV, as we
explain in the following paragraph.)  First the competing endoergic
channel
\begin{equation}
\mathrm{O} + \mathrm{H_3^+} \rightarrow \mathrm{OH} + \mathrm{H_2^+} - 1.77 \mathrm{eV} 
\label{competition1}
\end{equation}
appears to open up, resulting in a reduction in the measured
merged-beams rate coefficient.  This is followed at $\sim 3$~eV by the
opening up of another channel for forming OH$^+$, namely Reaction
(\ref{OH+other}), causing a compensating increase in the signal.  At
higher energies, three additional competing channels open up:
\begin{eqnarray}
\mathrm{O} + \mathrm{H_3^+} 
&\rightarrow& \mathrm{O} + \mathrm{H^+} + \mathrm{H_2} - 4.34 \mathrm{eV}, \\
&\rightarrow& \mathrm{O^+} + \mathrm{H} + \mathrm{H_2} - 4.36 \mathrm{eV}, \\
&\rightarrow& \mathrm{O} + \mathrm{H} + \mathrm{H_2^+} - 6.16 \mathrm{eV}. 
\label{competition2}
\end{eqnarray}
These are followed by a dramatic drop with increasing energy, which we
attribute to the competing endoergic reaction pathways taking up all
the flux of the reaction.

The shift in these structures to energies lower than the known
thresholds most likely comes from the molecular nature of the H$_3^+$,
with its range of possible ro-vibrational levels contributing to the
process \citep[as explained in more detail in][]{OCon14a}. This leads
to a smearing out of the thresholds with the relative energy, blurring
them together for the above-listed endoergic reactions. So we take the
highest measured value of the merged-beams rate coefficient, at a
relative energy of 0.89 eV, as the opening up of
Reaction~\ref{competition1}, which would be 1.77~eV for cold
H$_3^+$. The lower-than-expected energy in the initial decrease of the
merged-beams rate coefficient is probably due to the energy available
from the internal excitation of H$_3^+$ in our experiment. The $\sim$
0.87~eV energy difference lies within the $\sim 0.5-1$~eV range
inferred from theoretical calculations \citep{Anic84a} and
photodissociation measurements (X. Urbain, private communication).
Using the H$_3^+$ molecular partition function of \citet{Kyla11a} and
specifically their Equation~(8), this energy difference gives an
internal temperature of $\sim 3000$~K. This is slightly higher, but
still consistent, with the H$_3^+$ internal excitation of $\sim 2550$
inferred in our ${\rm C + H_3^+}$ work which used the same H$_3^+$ ion
source \citep{OCon14a}.
	
\subsubsection{\em O + H$_3^+ \rightarrow$ H$_2$O$^+$ + H }
	
Our merged-beams rate coefficient for Reaction~(\ref{o2h+}) decreases
with increasing relative energy. \citet{Bett99} describe the formation
of H$_2$O$^+$ as a two-step process. First a complex is formed where
the oxygen atom extracts H$^+$ from H$_3^+$ to form an OH$^+$ ion
adjacent to the remaining H$_2$. Then the OH$^+$ and H$_2$ undergo a
relative rotation after which the OH$^+$ extracts a hydrogen atom from
the H$_2$. This entire process takes time and could explain the
decrease seen in the rate coefficient. As the relative energy
increases, the time available for the reaction also decreases. Clearly
though, further theoretical and experimental studies are needed to better
understand the process.

At $\sim 0.1$ eV, we observe a strong decrease in the rate
coefficient, similar to that of Reaction (\ref{ch2+}) seen by
\citet{OCon14a}. The endoergic Reactions (\ref{OH+other}) -
(\ref{competition2}) are the probable explanation for this drop. The
shift to a lower relative energy is also likely due to the internal
excitation of H$_3^+$.

\subsection{Converting Translational Temperature to Thermal Rate Coefficients} 
\label{sec:thermal}

In order to convert our translational temperature rate coefficients
into a thermal rate coefficient, we follow the approach outlined in
\citet{OCon14a}. This method enables us to correct our results for
statistically populated fine-structure levels in the atomic O into
data relevant for thermally populated fine-structure levels. In this,
we follow the approximations made by Klippenstein et al. (2010), in
their calculations for the ${\rm O + H_3^+}$ collision system, namely
that the reaction proceeds adiabatically and that surface crossings
and intersystem transitions are both unimportant.

Another assumption that we make is that the internal excitation of the
H$_3^+$ does not affect the reaction. Theoretically, this approach is
supported by the long-range character of the entrance PES being
dictated by the polarization of the fine-structure state of the oxygen
in the field of a point charge \citep{Gentry}.  Experimentally, it is
supported by the good agreement between our ${\rm C + H_3^+}$ results
on internally excited H$_3^+$ and the mass-scaled results of Savi\'c
et al. (2005) for C on D$_3^+$ with an internal temperature of
77~K. It is also supported by the good agreement that we find below
between our thermal ${\rm O + H_3^+}$ results and those of
\citet{Fehsenfeld} and \citet{Mill00a} for H$_3^+$ with an internal
termperature of 300~K.

\begin{figure}[!p]
\centering
\includegraphics[width=1\textwidth]{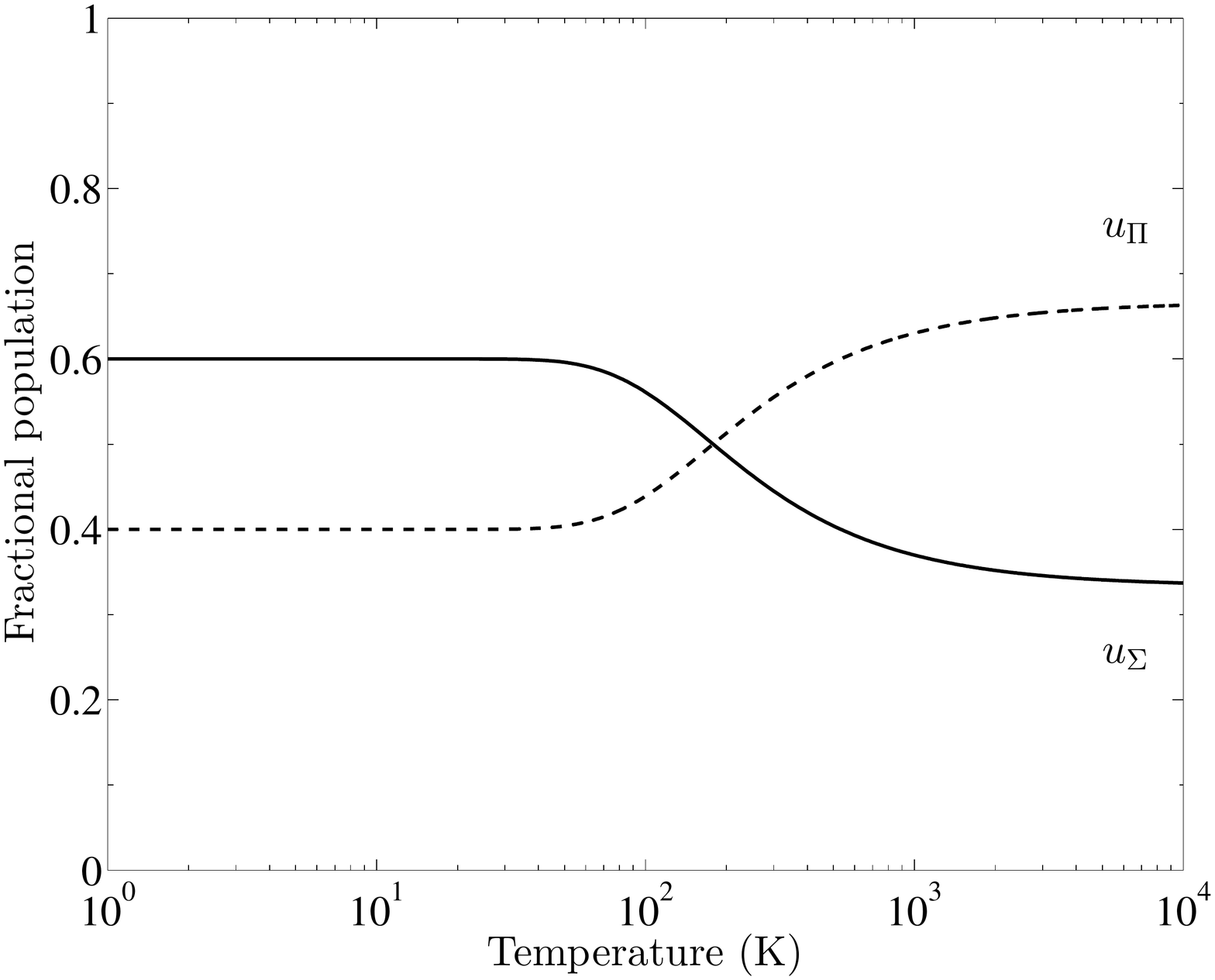}
\caption{Fractional population of the H$_3$O$^+$ attractive $^3\Sigma$
  and repulsive $^3\Pi$ symmetries vs.\ temperature.}
\label{Fig:surfacelevels}
\end{figure}

Taken all together, these approximations enable us to derive
temperature-dependent multiplicative scaling factors to convert our
translational temperature results into thermal results. For this we
use the theoretical study of \citet{Gentry} for ground term atomic
oxygen in the presence of a positive charge. \citet{Bett01a} and
\citet{Klip10a} also based their calculations on \citet{Gentry}, which
found that the nine states in the O($^3P$) manifold are split into
three attractive and six repulsive surfaces at long-range separation
of the reactants. The $M_J=0$ and $\pm1$ components of the five-fold
degenerate ground state $^3P_2$ state correlate with the attractive
$^3\Sigma$ PES, while the $M_J=\pm2$ components of the $^3P_2$ and all
components of $^3P_1$ and $^3P_0$, correlate with the repulsive
$^3\Pi$ surface. The partition functions for the attractive $^3\Sigma$
and non-reactive $^3\Pi$ PES shown in Figure \ref{Fig:surfacelevels}
are given by
\begin{eqnarray}
u_\Sigma & = & \frac{3}{5} u_2 \\
u_\Pi & = & u_0 + u_1 +\frac{2}{5} u_2
\end{eqnarray}
with $u_J$ defined by Equation (\ref{Jpartfunction}).

\begin{figure}[!t]
\begin{center}	
\includegraphics[width=1\textwidth]{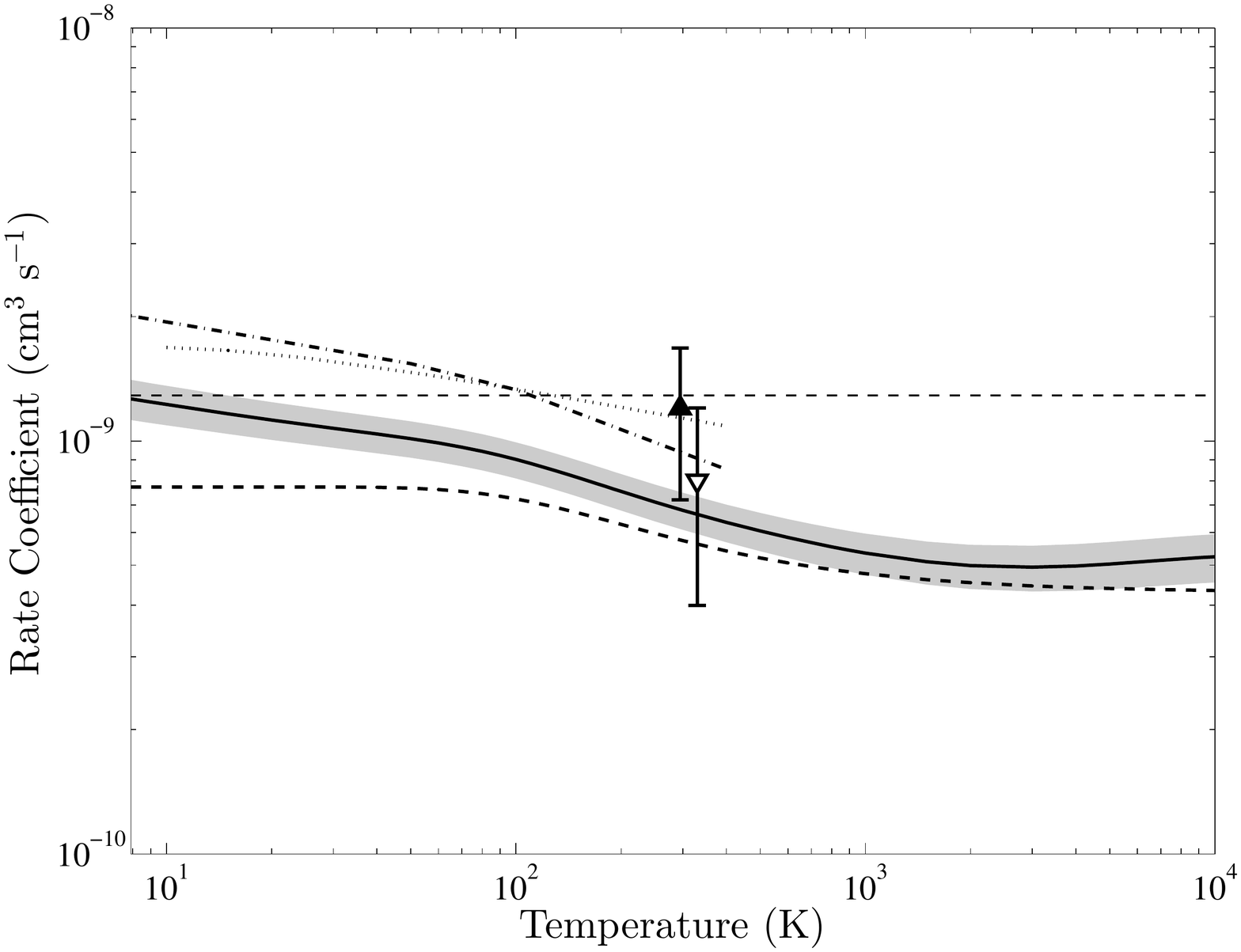} 
\caption{Our experimentally derived thermal rate coefficient
  vs. temperature for the sum of Reactions~(\ref{oh+}) and
  (\ref{o2h+}) is shown by the black solid line. The dashed lines are
  the Langevin rate coefficient unmodified (thin) and modified
  (thick). The dotted-dashed lines present the total theoretical
  calculations of \citet{Bett99}. The \citet{Klip10a} calculations are
  shown by the dotted line. The inverted open triangle is the 300~K
  experimental result of \citet{Fehsenfeld} shifted to 330~K for
  clarity, and the filled triangle the total 295~K experimental result
  of \cite{Mill00a}.}
\label{Fig:thermalratesum}
\end{center}
\end{figure}

The theoretical approach outlined above provides information only for
the total reaction probability, i.e., the sum of Reactions (\ref{oh+})
and (\ref{o2h+}). We are aware of only one published experimental
study for the branching ratios for forming OH$^+$ and H$_2$O$^+$,
namely the results of \citet{Mill00a} at 300~K. How those results
scale with temperature is uncertain. Indeed, \citet{OCon14a} found a
significant temperature dependence for the branching ratio in the
analogous ${\rm C + H_3^+}$ reaction. Lacking the necessary
temperature-dependent information on the branching ratios, in order to
convert our translational temperature results to a thermal rate
coefficient, we sum our translational temperature rate coefficients
for Reactions (\ref{oh+}) and (\ref{o2h+}) and then multiply the
result by 3$u_\Sigma$. This factor of 3 takes into account the fact
that in our measurement only one-third of the O fine-structure levels
contribute to the reaction process. The resulting summed thermal rate
coefficient is shown in Figure~\ref{Fig:thermalratesum}.

We can also use this theoretical approach to generate a
temperature-dependent Langevin rate coefficient. The classical
Langevin rate coefficient of $1.3 \times 10^{-9}$ cm$^3$~s$^{-1}$
\citep{Mill00a}, considers all of the O + H$_3^+$ symmetries involved
in the reaction process to be attractive. We can convert this value to
a temperature-dependent thermal rate coefficient by multiplying it by
the partition function $u_\Sigma$. Both the unmodified and modified
Langevin rate coefficient are shown in
Figure~\ref{Fig:thermalratesum}.
	
\subsection{Thermal Rate Coefficients}
\label{sec:thermal2}
	
Our experimentally derived thermal rate coefficient for ${\rm O +
  H_3^+}$ forming either OH$^+$ or H$_2$O$^+$ is in good agreement
with the unmodified Langevin rate coefficient at 10~K. At this
temperature, the modified Langevin rate coefficient lies a factor of
$\approx$ 1.6 below our experimental results. This is not surprising
since the charge-quadrupole interaction responsible for the growth of
the rate coefficient at low temperatures is absent from the Langevin
model.  As the temperature increases, the unmodified rate coefficient
becomes increasingly discrepant with the experimental results, growing
to a factor of $\approx$ 2 times larger at 300~K. Conversely the
modified rate coefficient comes into reasonable agreement with our
results, lying just outside the experimental error bars, but matching
the general temperature dependence.

The semi-classical results of \citet{Bett99} and those of
\citet{Klip10a} both shown in Figure~\ref{Fig:thermalratesum}, each
display a trend similar to our experimentally derived thermal rate
coefficient. However, both sets of values lie outside our experimental
error bars. The Bettens \& Collins results are a factor of $\approx$
1.7 larger at 10~K and a factor of $\approx$ 1.3 at 300~K, while the
\citet{Klip10a} results are a factor of $\approx$ 1.4 and 1.7 larger
at 10 and 300~K, respectively. The cause for the discrepancy is not
clear.  For the analogous summed thermal rate coefficient for ${\rm C
  + H_3^+}$, reasonable agreement was found between our results and
the calculations of \citet{Bett98a,Bett01a}.
	
Our derived thermal rate coefficient is in good agreement with the
experimental results from \citet{Fehsenfeld} at 300~K. In his flowing
afterglow experiment, he was unable to distinguish the products of the
reaction and thus gave only an overall thermal rate coefficient for O
on H$_3^+$. \citet{Mill00a} measured the rate coefficients of
Reactions (\ref{oh+}) and (\ref{o2h+}) at 295~K. We compare with their
summed results and find that their error bar overlaps with ours. So
even though their results are a factor of $\sim 1.8$ times larger than
ours, the agreement seems reasonable enough, given the challenge of
monitoring the atomic oxygen density in the flowing afterglow method.
	
Comparing all of the experimental thermal results for O on H$_3^+$ at
$\approx 300$~K, we find good agreement between our work with hot
H$_3^+$ ($\sim$ 3000~K) and the two published measurements on cold
H$_3^+$ ($\approx 300$~K).  Similarly good agreement was found at a
temperature of $\sim 1000$~K between our ${\rm C + H_3^+}$ work
compared to the mass-scaled results of \citet{Savi05a} for ${\rm C +
  D_3^+}$, which had an internal temperature of 77~K.  Taken together,
these experimental results strongly suggest that for reactions of the
type ${\rm X + H_3^+}$, the internal excitation of the H$_3^+$ does
not play a significant role in determining the total rate coefficient
for reacting, summed over all possible outgoing channels.  We
hypothesize that this behavior is due, in part, to the reaction being
driven primarily by the long range portion of the PESs for the
reaction systems, where the internal excitation of the molecule is not
yet felt.  Furthermore, H$_3^+$ has only a single bound electronic
symmetry and so the entire population has the same spin symmetry.
Thus any spin conservation in the reaction affects the entire H$_3^+$
population the same way, as opposed to a molecule with a population 
divided between two or more different spin multiplicities.  Clearly,
though, further experimental and theoretical work is needed to more
solidly understand the physics of the reaction.

\section{Astrochemical Implications} 
\label{astro}

We have used the gas-phase astrochemical code Nahoon \citep{Wake12a}
and the 2014 version of the KInetic Database for Astrochemistry
\citep[KIDA;][]{Wake15a} in order to investigate some of the
astrophysical implications of our findings, particularly for the
gas-phase chemistry of dense molecular clouds.  Though a complete
study is beyond the scope of this paper (for example, we leave for
future study the effects of surface or ice chemistry as well as any
depletion effects), these preliminary models do give some insight into
the impact of our results.  Following \citet{Wake15a}, the specific
cloud parameters used were a hydrogen nuclei density of $n_{\mathrm H}
= 2 \times 10^4$~cm$^{-3}$, a CRIR of $\zeta = 10^{-17}$~s$^{-1}$, and
a visual extinction of $A_{\rm V} = 30$.  We also adopted the initial
abundances given in their paper.

For these simulations we have taken our experimentally derived thermal
rate coefficient for ${\mathrm O} + {\mathrm H}_3^+$ for forming
either OH$^+$ or H$_2$O$^+$ and implemented it into Nahoon/KIDA.
Since the available fitting functions within Nahoon/KIDA were unable
to accurately fit our thermal rate coeffiient, we have fit it using
the recommended function of \citet{Novo13a}, namely
\begin{equation}
k(T) =
A\,\left(\frac{300 ~\rm K}{T}\right)^n + T^{-3/2}\sum_{i=1}^4 c_i \exp(-T_i/T).
\label{eq:plasmafitnew}
\end{equation}
The resulting fit parameters are given in Table~\ref{tab:plasmares}.
The deviations of the fitted $k$ from the data are less than
$\sim 0.5\%$ over the reported $1-10^4$~K temperature range.

We have implemented our thermal rate coefficient into Nahoon/KIDA, modifying
Nahoon to handle the above equation.  Since we do not know the
branching ratios for the formation of OH$^+$ and H$_2$O$^+$, we have
run simulations first assuming a branching ratio of 100\% for forming
OH$^+$ and later assuming 100\% for forming H$_2$O$^+$.  This is
justified as both ions are predicted to be highly transitory, rapidly
reacting with the abundant H$_2$ in the cloud to form H$_3$O$^+$ via
hydrogen abstraction, 
\begin{equation}
{\rm OH^+ \stackrel{H_2}{\longrightarrow} H_2O^+ 
\stackrel{H_2}{\longrightarrow} H_3O^+}.
\label{eq:CH+_Hab}
\end{equation}  
As we expected, we found no significant difference in our model
results for either branching ratio assumption.

\subsection{Formation of Water}

\begin{figure}[!t]
\begin{center}	
\includegraphics[angle=90,width=1\textwidth]{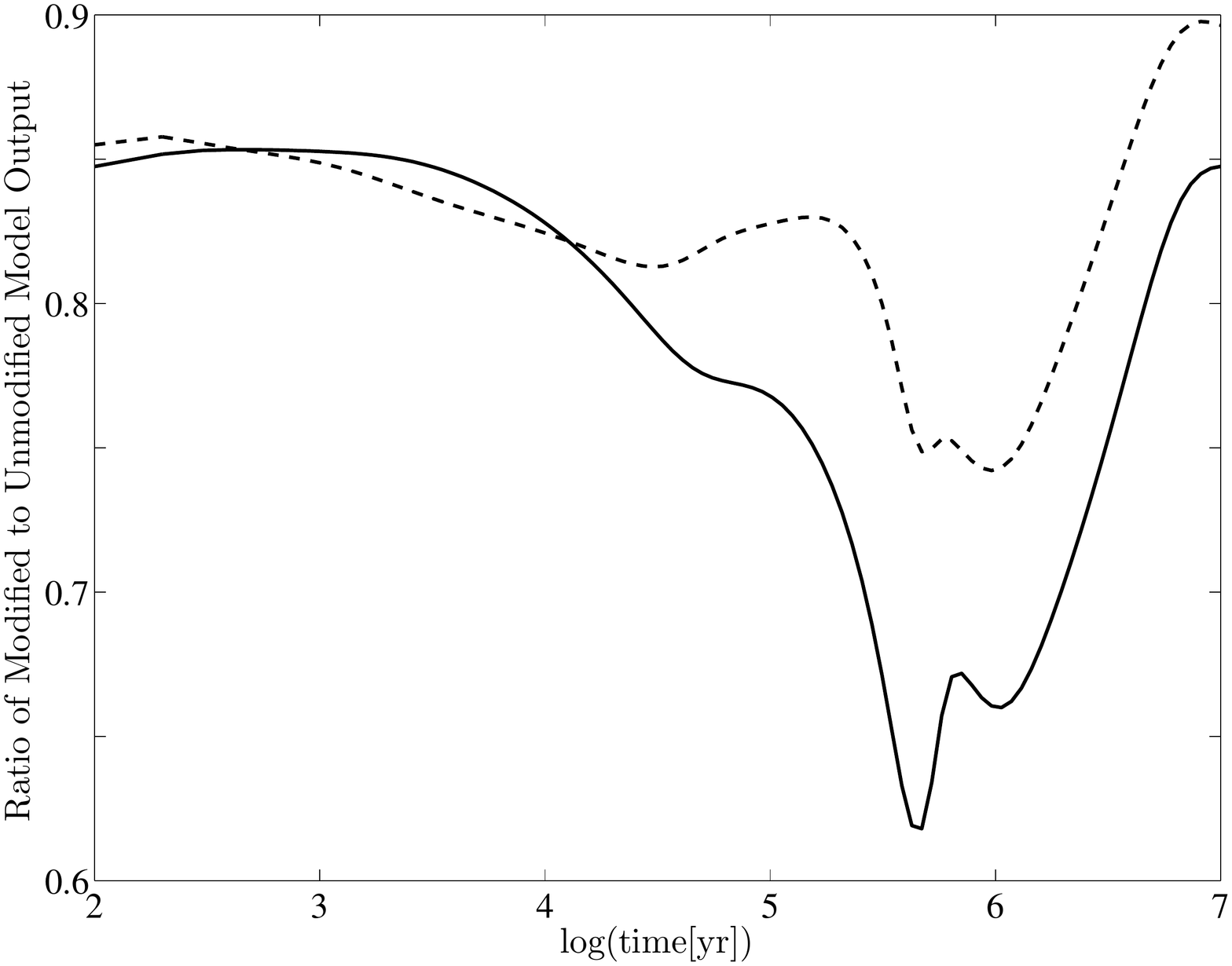} 
\caption{Ratio of the modified to unmodified predicted abundances for
  H$_3$O$^+$ (dashed line) and H$_2$O (solid line).  The ratio gives
  the modified Nahoon/KIDA results using our experimentally derived
  total rate coefficient for the ${\rm O + H_3^+}$ reaction divided by
  the unmodified Nahoon/KIDA results, which uses the rate coefficient
  from \citet{Klip10a}.}
\label{Fig:AbundanceRatios}
\end{center}
\end{figure}

To investigate the impact of our results on the formation of water, we
have calculated the H$_3$O$^+$ and H$_2$O abundances for a dense
molecular cloud at a temperature of 10~K.  Over most of the lifetime
of a cloud, H$_3$O$^+$ is the dominant gas-phase precursor of water,
forming H$_2$O via dissociative recombination
\begin{equation}
e^- + \mathrm{H_3O}^+ 
\to \mathrm{H_2O} + \mathrm{H}.
\label{eq:H3O+DR}
\end{equation}
Figure~\ref{Fig:AbundanceRatios} shows the ratios of the predicted
abundances using our results relative to those from the unmodified
Nahoon/KIDA, which uses the theoretical rate coefficient results of
\citet{Klip10a}.  

At a cloud age of 10$^2$ years, the abundance of both H$_3$O$^+$ and
H$_2$O are reduced by $\approx 15$\%.  These changes can be traced
back to our reduced rate coefficient for forming OH$^+$ or H$_2$O$^+$,
which go on to H$_3$O$^+$ via hydrogen abstraction and then water via
dissociative recombination.  Naively, one might expect the H$_3$O$^+$
and H$_2$O abundances to be reduced by the ratio of our rate
coefficient relative to that of \citet{Klip10a}, which is $\approx
0.72$ at 10~K.  However, our lower rate coefficient results in a
reduced destruction rate for the H$_3^+$ and thus a higher predicted
abundance, giving an abundance ratio at 10$^2$ years of $\approx 1.18$
compared to the unmodified Nahoon/KIDA results.  Multiplying these two
factors together explains the $\approx 15$\% reduction seen at short
times.

At longer times, the observed reductions in H$_3$O$^+$ and H$_2$O
become much more significant.  Identifying the cause for this, though,
is complicated, due to an increase in the complexity of the relevant
chemical network.  As an example, we discuss here the intermediate
times between $\sim 10^{3.5}$ and $10^{5.9}$ years.  During this time,
along with reaction~(\ref{eq:H3O+DR}), two additional important
channels for forming water open up, namely
\begin{equation}
\mathrm{C_3} + \mathrm{H_3O^+} 
\to \mathrm{H_2O} + \mathrm{C_3H^+}
\end{equation}
and
\begin{equation}
\mathrm{HCN} + \mathrm{H_3O^+} 
\to \mathrm{H_2O} + \mathrm{HCNH^+}.
\end{equation}
Indeed, these two channels are predicted to dominate over dissociative
recombination for part of this epoch.  At longer times, dissociative
recombination again comes to dominate.

New channels for forming H$_3$O$^+$ also open up between $\sim
10^{4.5}$ and $10^{5.8}$ years, including
\begin{equation}
\mathrm{ H}_2\mathrm{O} + \mathrm{H}_3^+ 
\to \mathrm{H}_2 + \mathrm{H}_3\mathrm{O}^+
\end{equation}
and
\begin{equation}
\mathrm{H_2O + HCO^+} 
\to \mathrm{CO + H_3O^+}.
\label{eq:H2O+HCO+}
\end{equation}
However, these reactions lead to destruction of H$_2$O as well, since
only a fraction of the resulting H$_3$O$^+$ gets converted back into
water.  We find that with the modified chemistry, the larger decrease
seen in the relative H$_2$O abundance (compared to that for
H$_3$O$^+$) can be traced back, in part, to the increased H$_3^+$
abundance during this epoch, leading to an enhancement in HCO$^+$
formation via
\begin{equation}
{\rm CO + H_3^+}
\to {\rm HCO^+ + H_2}.
\end{equation}
The resulting HCO$^+$ then leads to an increased reduction in the
water abundance via reaction~(\ref{eq:H2O+HCO+}) for the modified
chemistry.  

\subsection{Cosmic Ray Ionization Rate}

In dense molecular clouds, the CRIR $\zeta$ can be constrained through
observations of the H$_3^+$ column density
\citep[e.g.,][]{McCa99a,Klip10a,Oka13a}.  The H$_3^+$ is formed in a
two-step process, beginning with the ionization of H$_2$ by cosmic
rays; other ionization processes of H$_2$ are insignificant.  At
typical dense cloud densities of $n_{\rm H} \sim 10^4$~cm$^{-3}$, the
resulting H$_2^+$ rapidly reacts with a neutral H$_2$ molecule, in
less than a day \citep{Oka13a}, meaning that H$_3^+$ generation is
completely determined by the CRIR, with a formation rate versus time
$t$ of
\begin{equation}
\left[{{\rm d}n_{\rm H_3^+} \over {\rm d}t}\right]_{\rm form}
=
\zeta n_{\rm H_2},
\label{eq:form}
\end{equation}
where $n_{\rm H_2}$ is the number density of molecular hydrogen.  The
dominant destruction mechanisms of H$_3^+$ in dense clouds are binary
chemical reactions with other species, primarily neutral since the
electron abundance is too low for dissociative recombination to play a
significant role.  Thus we can write the destruction rate as
\begin{equation}
\left[{{\rm d}n_{\rm H_3^+} \over {\rm d}t}\right]_{\rm dest}
=
\sum_i k_i n_i n_{\rm H_3^+},
\label{eq:dest}
\end{equation}
where $k_i$ is the thermal rate coefficient for the reaction of species 
$i$ with H$_3^+$ and $n_i$ is the number density of species $i$.  

In the quasi-equilibrium of dense molecular clouds, we can equate 
Equations~(\ref{eq:form}) and (\ref{eq:dest}) giving
\begin{equation}
\zeta n_{\rm H_2} 
=
\sum_i k_i n_i n_{\rm H_3^+}.
\label{eq:equilibrium}
\end{equation}
As described in \citet{McCa99a}, \citet{Klip10a}, and \citet{Oka13a}, 
the properties of dense clouds enable us to make the approximation
\begin{equation}
N_{\rm H_3^+} = L n_{\rm H_3^+},
\label{eq:approximation}
\end{equation}
where $N_{\rm H_3^+}$ is the observed column density of H$_3^+$ and
$L$ the absorption path length through the cloud.  Combining
Equations~(\ref{eq:equilibrium}) and (\ref{eq:approximation}) and
rearranging gives
\begin{equation}
\zeta L
=
\left({N_{\rm H_3^+} \over n_{\rm H_2}}\right) \sum_i k_i n_i.
\label{eq:zetaL}
\end{equation}

\begin{figure}[!t]
\begin{center}	
\includegraphics[width=1\textwidth]{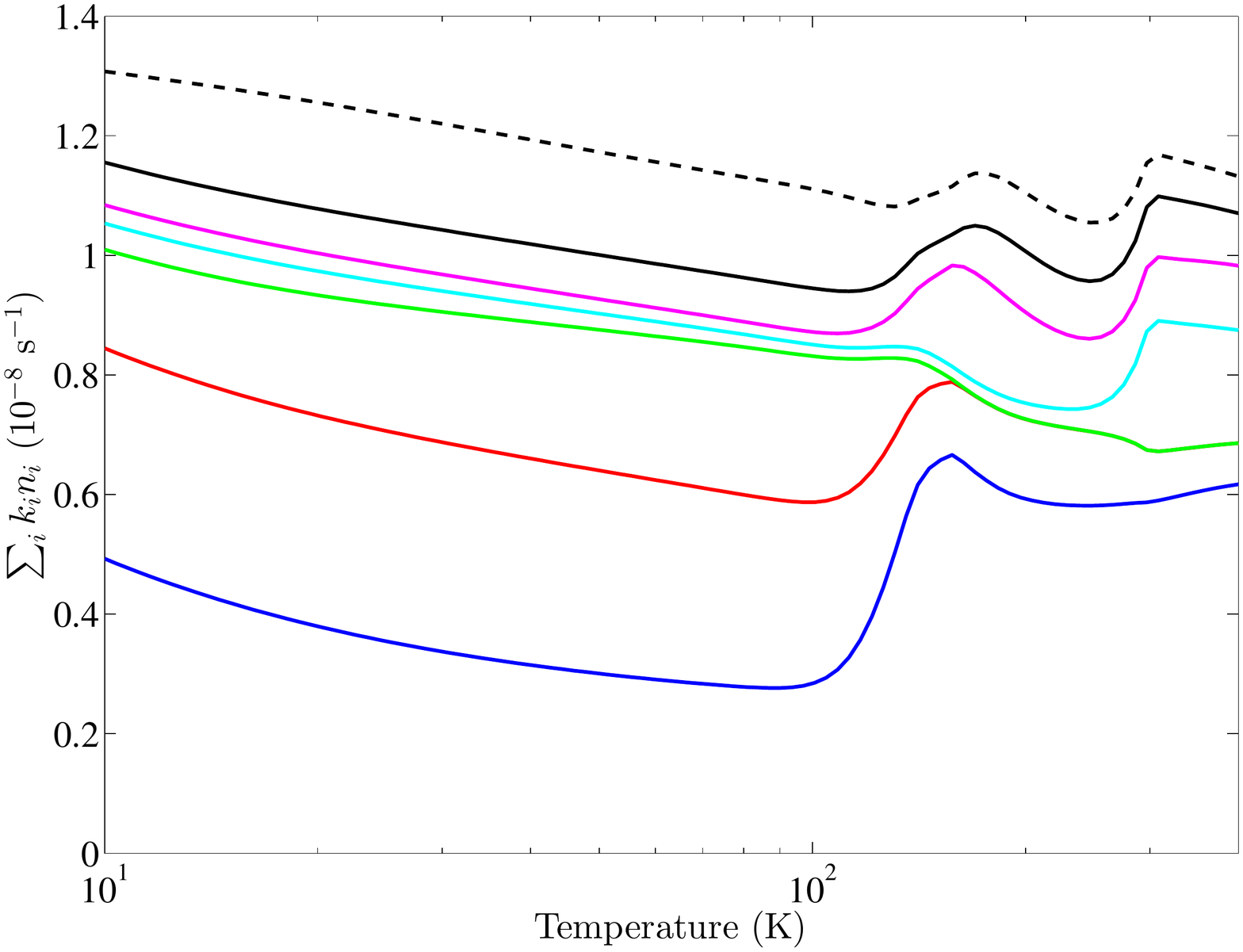} 
\caption{Predicted factor $\sum_i k_i n_i$ at a cloud age of
  $10^5$~years using the modified and unmodified Nahoon/KIDA.  The
  solid colored lines correspond to the number of terms being included
  in the summation for the modified factor, starting with dark blue
  for just CO, red for the addition of O, green for C, light blue for
  H$_2$O, violet for HCN/HNC, and black for the addition of all other
  species.  The predicted factor for all reactants using the
  unmodified Nahoon/KIDA is shown by the black dashed curve.}
\label{Fig:1e5}
\end{center}
\end{figure}

\begin{figure}[!t]
\begin{center}	
\includegraphics[width=1\textwidth]{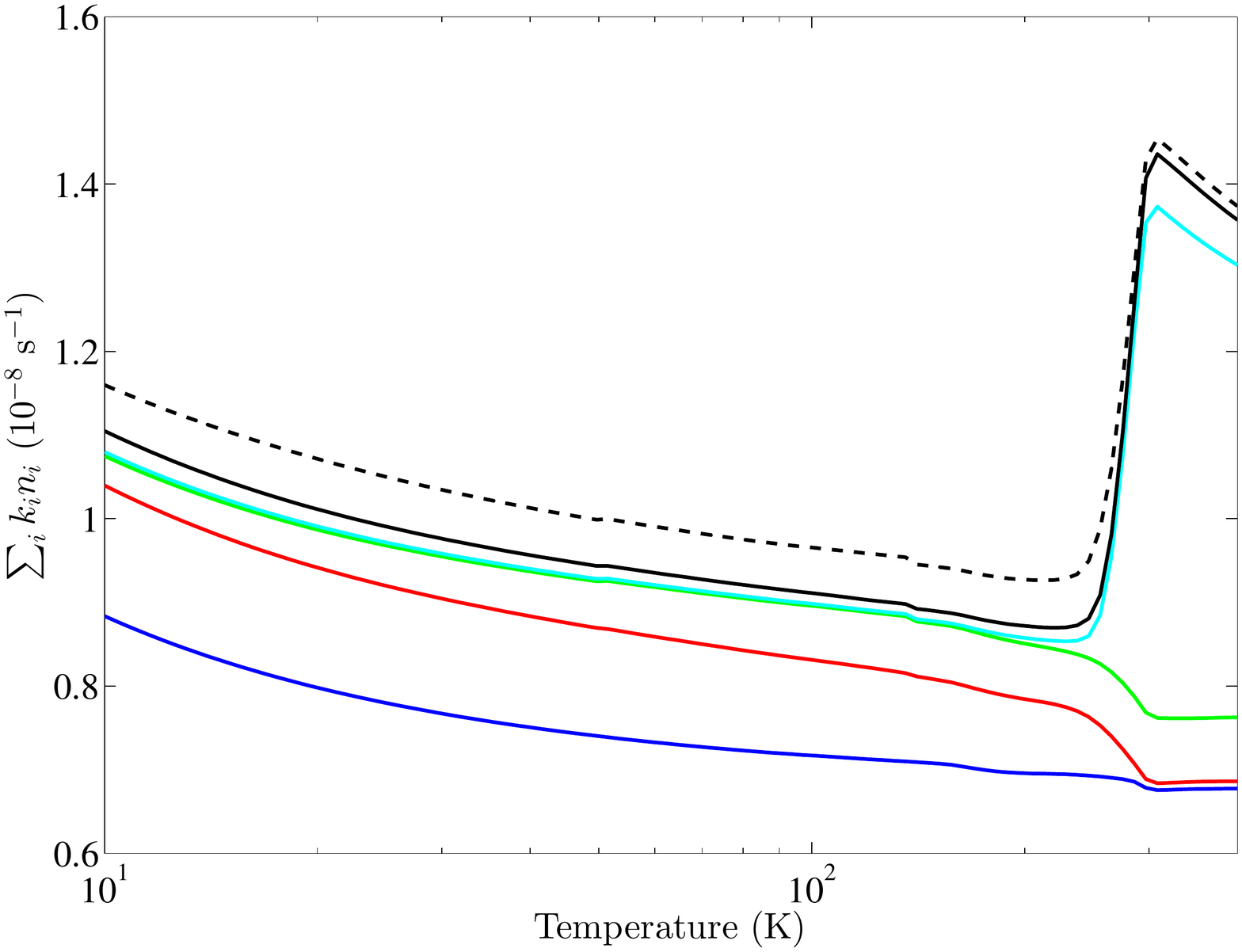}
\caption{Same as Fig.~\ref{Fig:1e5} but for a cloud age of $10^6$~years.
  Here the cumulative additions of the reactants $i=\{\rm CO, O, N_2,
  H_2O, all\ other\ species\}$ are shown by the dark blue, red, green,
  light blue, and black solid lines, respectively.}
\label{Fig:1e6}
\end{center}
\end{figure}

The quantity $\zeta L$ can thus be constrained by a combination of the
observed H$_3^+$ column density and $n_{\rm H_2}$ number density and
astrochemical calculations of the individual terms contributing to the
summation over species.  Previous studies into
Equation~(\ref{eq:equilibrium}) have limited the number of species
considered.  For example, \citet{McCa99a} considered only reactions
with CO, \citet{Klip10a} also included O, and \citet{Oka13a} added to
these N$_2$ and electrons.  Here we have used the modified Nahoon/KIDA
to calculate $\sum_i k_in_i$ for the most significant species as well
as for all species.  For comparison, we have also calculated the
unmodified Nahoon/KIDA results for all species.  Figures~\ref{Fig:1e5}
and \ref{Fig:1e6}, respectively, show our results for cloud ages of
$10^5$ and $10^6$~years over the temperature range of $10-400$~K.
These times lie within the range of commonly inferred cloud ages.

At $10^5$~years and below $\sim 160$~K, the most important reactants
with H$_3^+$ are CO, O, C, H$_2$O, and the HCN/HNC isomers, in
descending order of importance.  The contributions of each of these
species are shown by the various colored solid lines in
Fig.~\ref{Fig:1e5} and the total $\sum_i k_in_i$ due to all species by
the solid black line.  Above $\sim 250$~K, the abundance of C
decreases dramatically and reactions with C become unimportant.  It is
also interesting to note that the structure due to CO largely gets
washed out by the contributions of O, C, and H$_2$O.  Additionally,
the structures seen in the total $\sum_i k_in_i$ are due primarily to
the HCN/HNC isomers between $\sim 120$ and 250~K and due to water
above 250~K.  The increase due to H$_2$O is a result of the higher
temperatures enabling neutral-neutral reactions leading to water to
become important.  Using the unmodified Nahoon/KIDA yields a total
summation shown by the black dashed line, which is up to $\sim 15\%$
larger than the modified value.

At $10^6$~years and below $\sim 240$~K the most important reactants
with H$_3^+$ are CO, O, and N$_2$, in decreasing order of importance.
The contributions of each of these species are shown by the various
colored solid lines in Fig.~\ref{Fig:1e6} and the total $\sum_i
k_in_i$ due to all species by the solid black line.  Above $\sim
240$~K, the abundance of O decreases dramatically and reactions with O
become unimportant.  Conversely, the abundance of H$_2$O increases as
neutral-neutral reactions leading to water become important.  Using
the unmodified Nahoon/KIDA yields a total summation shown by the black
dashed line, which is up to $\sim 8\%$ larger than the modified value.

These findings indicate that constraints to the value $\zeta L$
depend, in part, on using a complete chemical model, knowing the age
and temperature of the observed cloud, and using accurate reaction
rate coefficients.  The calculated abundances $n_i$ also depend on the
adopted values for $\zeta$ and $n_{\rm H_2}$ in the astrochemical
model, implying that one will have to iterate the model in order to
achieve convergence with the observations for the quantity $\zeta L$.
Our experimentally derived thermal rate coefficient reported here
helps to improve the reliability of this approach to determining
$\zeta L$, but it is clear that there are many additional parameters
which also need to be well constrained.

\section{Summary} 
\label{summary}

Using a novel merged fast-beams apparatus, we have measured the cross
sections for ${\rm O + H_3^+}$ forming OH$^+$ and H$_2$O$^+$.  Our
measurements were performed for statistically populated O($^3P_J$) in
the ground term and internally hot H$_3^+$ ($\sim 2500-3000$~K).
Using state-of-the-art theory as a guide to account for the
temperature dependence of the O fine-structure $J$-levels, we have
converted our results into a thermal rate coefficient for forming
either OH$^+$ or H$_2$O$^+$.  The good agreement that we find with the
two published flowing afterglow meaurements at a temperature of
$\approx 300$~K (and a corresponding level of H$_3^+$ internal
excitation) strongly suggests that the H$_3^+$ internal excitation
does not significantly affect the thermal rate coefficient for this
reaction.  The Langevin value is in good agreement with our results at
10~K but a factor of 2 higher at 300~K.  The two published
semiclassical results lie a factor of $\sim 1.5$ above our results
over this temperature range.  We have implemented our results into the
astrochemical code Nahoon to explore some of the astrophysical
implications of our results.  For example, for dense clouds at 10~K,
we find a reduction in the predicted water abundance by up to nearly
40\% at certain times in the lifetime of the cloud.


	
\acknowledgments

The authors thank V. Wakelam for stimulating conversations.  This work
was supported in part by the Advanced Technologies and Instrumentation
Program and the Astronomy and Astrophysics Grants Program in the NSF
Division of Astronomical Sciences. X.U. is Senior Research Associate
of the FRS-FNRS.  S.V. was supported in part by the NSF Research
Experience for Undergraduates program.
	
		
\clearpage
\bibliography{OH3plus}

\clearpage

	\clearpage

	\clearpage

	\clearpage

	\clearpage

\clearpage

\clearpage
	
\clearpage
	
\clearpage
			
\clearpage
			
\clearpage

\clearpage

\begin{table}	
\begin{center}
\caption{Typical values of the statistical-like uncertainties for Equation~(\ref{rateeq}) for a single data~run.}   
\label{rateval}
\begin{tabular}{lccc}
\hline
\hline
Source & Symbol & Value & Uncertainty (\%) \\
\hline 
Signal rate      & $S$                       & 1-15 Hz                          & $\leq$ 4\\
O velocity       & $v_{\mathrm n}$           & $5.8 \times 10^7$~cm~s$^{-1}$ 	& $\ll$1   \\
H$_3^+$ velocity & $v_{\mathrm i}$	  		 & $5.8 \times 10^7$~cm~s$^{-1}$ 	& $\ll$1   \\
O current        & $I_{\mathrm n}$           & 23 nA                            & 5        \\
H$_3^+$ current	 & $I_{\mathrm i}$	         & 225 nA                           & 5        \\
Overlap factor   & $\left<\Omega(z)\right>$  & 2.7 cm$^{-2}$                    & 10       \\
\hline
Statistical-like uncertainty (single run) & & &  13 \\
\hline
\hline
\end{tabular}		
\end{center}	
{Note. - The confidence level for each value is taken as equivalent to
a $1\sigma$ confidence level.  The total statistical-like
uncertainty is calculated by treating each individual uncertainty as
a random sign error and adding all in quadrature.}
\end{table}

\clearpage
\begin{table}
\begin{center}
\caption{Same as Table 1 but for the systematic uncertainties for all data runs.} 
\label{ratevalsys}
\begin{tabular}{lccc}
\hline
\hline
Source & Symbol & Value & Uncertainty (\%) \\
\hline
Analyzer transmission                   & $T_{\mathrm a}$ 		& 0.74     & 3 \\
Grid transmission                       & $T_{\mathrm g}$ 		& 0.90     & 1 \\
Neutral transmission                    & $T_{\mathrm n}$ 		& 0.94     & 2 \\
Neutral detector calibration 			& $\gamma$ 			& 2.6      & 12 \\
CEM efficiency       					& $\eta$            & 0.99     & 3 \\
Interaction length    					& $L$               & 121.5 cm & 2 \\
\hline 
Total systematic uncertainty & & & 13\\
\hline
\hline
\end{tabular}
\end{center}
\end{table}
		
	\clearpage

	\begin{table}[!p]
	\begin{center}
	\caption{Fit parameters for the cross section of Reactions~(\ref{oh+}) and (\ref{o2h+}) in units of cm$^2$ for $E$ in eV, using Equation~(\ref{Eqn:CS}).}
	\label{tab:CS}
	\begin{tabular}{cccccc}
	\hline \hline
	Reaction        &   			&           	& Parameters	&       		&        \\
	\cline{2-6} 				
					& $a_0$			& $a_{1/2}$ 	& $b_{1}$ 		& $b_{2}$ 		& $b_{4}$ \\
	\hline
	(\ref{oh+})   	& 3.7314E-16 	& 2.1237E-16 	&	-			& 3.6414E-02	& 5.0532E-04 \\
	(\ref{o2h+})  	& 9.8531E-17 	&	-			& -4.0668E-02	& -4.1891	 	& 517.80	\\
	\hline \hline
	\end{tabular}
	\end{center}
	\end{table}

	\clearpage

	\begin{table}[!p]
	\begin{center}
	\caption{Fit parameters for the kinetic temperature rate coefficient for Reactions~(\ref{oh+}) and (\ref{o2h+}) in units of cm$^3$ s$^{-1}$ for $T$ in K, using Equation~(\ref{Eqn.thermalrate}).}
	\label{tab:thermal}
	\begin{tabular}{ccccccc}
	\hline \hline
	Reaction        &   			&           	&               & Parameters	&       		&        \\
	\cline{2-7}				
					& $a_0$			& $a_{1/2}$ 	& $a_{1}$ 		& $b_{1/2}$ 	& $b_{1}$ 		& $b_{3/2}$ \\
	\hline 
	(\ref{oh+})   	& 5.1142E-10	& 2.6568E-11	& 9.8503E-15	& 1.6747E-02	& -9.9613E-5	& 1.1006E-6 \\
	(\ref{o2h+})    & 4.2253E-10	& -				& -				& 3.4977E-03	& -1.4126E-04	& 6.3584E-05 \\

	\hline \hline
	\end{tabular}
	\end{center}
	\end{table}
	
\clearpage

\begin{deluxetable}{crll}
\tablecolumns{3} 
\tablewidth{0pc} 
\tablecaption{ Fit parameters using Equation~(\ref{eq:plasmafitnew}) 
for the thermal rate coefficient of ${\rm O + H_3^+}$ forming either
OH$^+$ or H$_2$O$^+$. 
		\label{tab:plasmares}
		}
\tablehead{ \colhead{Parameter} & \multicolumn{2}{c}{Value}  &
\colhead{Units}\\
\cline{2-3}
 & \colhead{$x$}& \colhead{$y$} & 
}
\startdata
$A$	& \phs $7.39$	& $-10$ \phn & ${\rm cm^3\ s^{-1}}$ \\
$n$	& \phs $1.46$	& \phn $-1$ & dimensionless\\
$c_1$	& $6.32$	& \phn $-8$ & ${\rm K^{3/2}\ cm^3\ s^{-1}}$ \\
$c_2$	& \phs $-3.00$  & \phn $-6$ & ${\rm K^{3/2}\ cm^3\ s^{-1}}$ \\
$c_3$	& $-1.17$	& \phn $-5$ & ${\rm K^{3/2}\ cm^3\ s^{-1}}$ \\
$c_4$	& $5.76$	& \phn $-4$ & ${\rm K^{3/2}\ cm^3\ s^{-1}}$ \\
$T_1$	& \phs $7.47$	& \phs\phn $1$ & K \\
$T_2$	& \phs $5.40$	& \phs\phn $2$ & K \\
$T_3$	& \phs $1.83$	& \phs\phn $3$ & K \\
$T_4$	& \phs $1.90$	& \phs\phn $4$ & K \\
\enddata
\tablecomments{\doublespace
The value for each parameter is given by $x \times 10^{y}$. }
\end{deluxetable}

\end{document}